\documentclass[12pt]{iopart}
\usepackage{graphicx}
\newcommand{\snoccfluxunc}{1.59^{+0.08}_{-0.07}\mbox{(stat)}^{+0.06}_{-0.08}\mbox{(syst)}} 
\newcommand{\snoesfluxunc}{2.21^{+0.31}_{-0.26}\mbox{(stat)}~\pm{0.10}~\mbox{(syst)}} 
\newcommand{\snoncfluxunc}{5.21\pm 0.27~\mbox{(stat)}~\pm0.38~\mbox{(syst)}} 

\begin{document}

\title{Solar Neutrino Measurements}
\author{A. B. McDonald}
\address{SNO Institute, Queen's University, Kingston, Canada K7L 3N6}

\begin{abstract}
A review of solar neutrino experiments is provided, including experimental measurements to date and proposed future measurements. Experiments to date have provided a clear determination that solar neutrinos are undergoing flavor transformation and that the dominant mechanism for this transformation is oscillation. The mixing parameters are well defined and limits are placed on sub-dominant modes. The measurements also provide strong confirmation of solar model calculations. New experiments under development will study neutrino oscillation parameters and sub-dominant modes with greater precision and will investigate solar fluxes further, concentrating primarily on the low energy pp, $^7$Be, pep and CNO reactions.   

\end{abstract}


\maketitle

\section {Introduction}

The study of solar neutrinos was initiated by the pioneering experimental efforts of Ray Davis coupled
with theoretical work of John Bahcall. The original objective was to study the physics of the Sun. The
set of measurements that have been carried out over more than 35 years have provided detailed information
on solar properties and have confirmed the accuracy of the Standard Solar Model. These measurements have also 
provided extensive information on the physics of neutrinos themselves,
including observations of flavor change that extend beyond the Standard Model of Elementary Particle Physics. The theoretical
aspects of solar neutrino measurements and their impact on solar models and neutrino properties are described in detail
in the following paper~\cite{BPG04} by Bahcall and Pena-Garay. The present paper will describe the experiments that have been carried
out to date and will provide a description of potential future experiments and their objectives.

\section{General description of solar neutrino measurements}

The models of the Sun have been developed to a high degree of detail over many years. The reactions that provide the thermonuclear power for the
Sun also emit copious numbers of neutrinos with energies ranging up to about 18 MeV. (See Figure 1 and Table 1 in
reference \cite{BPG04}). These neutrinos are very difficult to detect because of the very low neutrino cross sections 
and invariably require deep underground locations to get away from radioactive
backgrounds induced by cosmic rays.

The detection reactions that have been used for solar neutrinos are of two types:
\begin{itemize}
\item 
Radiochemical measurements involving the transformation of atoms of an element such as Cl or Ga into another
 radioactive element through inverse
 beta decay induced by electron neutrinos. These reactions can be observed by sweeping
 the radioactive elements from the detector volume and observing the subsequent decay.  The measurements on Chlorine 
by Ray Davis and his co-workers~\cite{davis1} and on Ga by the SAGE~\cite{SAGE}, GALLEX and GNO~\cite{GNO} experiments typically
 involve collection periods on the order of a month and subsequent decay measurement periods of many months.
\item 
The observation of events in real time produced by neutrino interactions in light water (Kamiokande~\cite{KAMIOKA} and Super-Kamiokande~\cite{SK})
and heavy water (SNO)~\cite{SNOsalt} via the Cerenkov process. The reactions
 that have been used involve elastic scattering of neutrinos from electrons, the inverse beta decay reaction 
on deuterium producing energetic electrons and inelastic scattering on deuterium producing free neutrons. The inverse
beta decay on deuterium is sensitive only to electron neutrinos, the inelastic scattering on deuterium is equally
 sensitive to all neutrino types and the elastic scattering on electrons is predominantly sensitive to electron neutrinos,
with a small (about 14\%) sensitivity to other neutrino types. 
\end{itemize}

These experiments have a variety of neutrino energy thresholds and are therefore sensitive to different combinations
of the neutrino-producing reactions in the Sun. Figure 1 of reference \cite{BPG04} shows the thresholds for the
above experiments and illustrates that the Gallium experiments have about one-half of their sensitivity from the pp reaction,
the Chlorine-based measurements have sensitivity primarily to the $^8$B neutrinos with some sensitivity to $^7$Be and the
light and heavy water-based experiments are sensitive almost exclusively to the $^8$B neutrinos with a negligible
contribution from the hep neutrinos.

Starting with the measurements on Chlorine and continuing with measurements on Gallium and
light water, too few neutrinos were observed in comparison to the predictions of solar models. This came to be known as the
Solar Neutrino Problem and was extensively discussed in the scientific literature. The attempts to explain the experimental
data generally involved a) changes to the solar models to modify the predicted neutrino fluxes or b) proposals for new physical
properties of neutrinos, such as oscillations among neutrino types leading to a smaller flux of electron neutrinos reaching
the detectors. The initial detectors were either exclusively or predominantly sensitive to electron neutrinos and so neutrino oscillation would result in lower observed fluxes. The oscillation of massive neutrinos can be modified by interactions with electrons in the Sun (or the Earth)
(the Mikheyev-Smirnov-Wolfenstein or MSW effect~\cite{MSW}) that can result in neutrino-energy-dependent probabilities for oscillation.
In addition, other mechanisms such as neutrino decay~\cite{NuDecay} or Resonant Spin-Flavor Precession (RSFP)~\cite{RSFP} were considered as explanations
for the small observed fluxes. Very detailed measurements of the energy dependence and time-dependence of $^8$B fluxes were carried out by
the Super-Kamiokande experiment with no statistically significant effects observed. Calculations of possible
 changes to solar models did not provide modifications to neutrino fluxes that agreed with the observed data. A firm conclusion on whether neutrino flavor change was taking place depended on solar model calculations of the relative size of the various neutrino fluxes. Although solar models provided very good agreement with all other solar properties, including helioseismology measurements~\cite{helio}, it was difficult to be certain whether changes to solar models or changes to neutrino properties might be required to explain the discrepancy between experiment and theory.

Subsequently, the Sudbury Neutrino Observatory (SNO)
experiment performed measurements of the interactions of $^8$B neutrinos with deuterium and was able to demonstrate
in a manner independent of detailed solar model flux calculations that neutrino flavor change was definitely taking place. The measurement showed that the flux of all active neutrinos was about three times the flux of electron neutrinos, clearly indicating the transformation of electron neutrinos to other active types. In addition, the measurement of the total active $^8$B neutrino flux was found to be in excellent agreement with the prediction of the Standard Solar Model Calculations.

The set of solar neutrino measurements to date defines a well-localized region of parameter space for neutrino mixing that also describes the oscillation of reactor anti-neutrinos observed recently by the KamLAND experiment. The observation of similar parameters for oscillation in these two cases provides strong restrictions on CPT, as well as restricting possible contributions from sterile neutrinos, RSFP, neutrino decay, and Flavor-Changing Neutral Current interactions. This overall set of measurements has thereby made a significant impact on our understanding of the Sun and of the basic properties of neutrinos. Neutrino physics has now moved to a precision phase and future measurements are aimed at the definition of neutrino properties and parameters of the Maki-Nakagawa-Sakahata-Pontecorvo (MNSP)~\cite{MNSP} mixing matrix with higher precision. They also will improve our knowledge of possible sub-dominant processes mentioned above. Future solar neutrino measurements are primarily aimed at observation of low energy fluxes and will thereby extend our detailed understanding of the neutrino processes in the Sun. The following sections will describe the measurements to date as well as further experiments that are being developed.

\section {Measurements to date}

\subsection{Radiochemical Detectors}
\subsubsection{Homestake Cl Experiment}

In the late 1960's, Ray Davis and
 co-workers \cite{davis1} began their pioneering solar neutrino measurements using 680 tons of liquid
 perchlorethylene sited 1480 meters underground (4300 meters water equivalent (m.w.e))in the Homestake
 gold mine near Lead, South Dakota, USA to observe electron neutrinos from the Sun. The electron neutrino capture reaction on $^{37}$Cl produces radioactive $^{37}$Ar atoms (35-day half life) in the liquid. The $^{37}$Ar atoms are flushed with helium gas every 100 days and condensed into sensitive proportional counters. About 0.5 argon atoms per day are produced. The capture reaction on $^{37}$Cl is calculated to be sensitive primarily to neutrinos from $^8$B decay in the Sun, but there is substantial sensitivity
 to neutrinos from $^7$Be as well.

 The detector is sited underground to reduce $^{37}$Ar production by cosmic rays. Detector and proportional counter
 materials are carefully selected to minimize contributions from background radioactivity. Background rates of $^{37}$Ar
 from sources other than the Sun are restricted to about 0.02 atoms per day. Non-radioactive tracer isotopes $^{36}$Ar
 and $^{38}$Ar are injected during successive runs to determine the argon extraction efficiency and are measured by mass
 spectrometry. Pulse shape analysis is used on the pulses from the proportional counter to reduce counter backgrounds
 from other than $^{37}$Ar decay.

The detector has been operated from 1968 until 2002 and the cumulative average of the data \cite{davis1} is
 $2.56 \pm 0.16(stat) \pm 0.16(syst)$ SNU, where a Solar Neutrino Unit (SNU) is defined as one electron neutrino capture
 per $10^{+36}$ atoms of $^{37}$Cl per second. This result is substantially smaller than the predictions of Standard
 Solar Model~\cite{BPG04} of $8.5 ^{+1.8}_{-1.8}$ SNU (Model BP04).

\subsubsection{Ga-Based Experiments: SAGE, GALLEX, GNO}

Experiments studying the electron neutrino capture reaction on Ga ($^{71}$Ga($\nu_e$,e$^-$)$^{71}$Ge)
are being carried out in the Baksan laboratory (depth of 4700 m.w.e.) in Russia (SAGE) \cite{SAGE} with about 60 tons of liquid Ga metal
 and in the Gran Sasso laboratory (3300 m.w.e) in Italy (GALLEX, GNO) \cite{GNO} with 30.3 tons of Ga in a
 concentrated GaCl$_{3}$-HCl solution.
 
The radioactive decay of $^{71}$Ge (half-life 11.4 days) is observed after extraction and deposition into low-background proportional counters. 
Somewhat different extraction techniques are used for the two experiments. In SAGE, the Ge
 is removed by extraction into an aqueous solution via an oxidation reaction, followed by concentration, conversion
 to GeCl$_{4}$ and synthesis to GeH$_{4}$ for use in the proportional counter. In GALLEX and GNO, volatile
 $^{71}$GeCl$_{4}$ is formed
 directly in the target solution, swept out by nitrogen gas, absorbed in water and then converted to GeH$_{4}$. 
 The detectors are very carefully made from low radioactivity materials and external background is carefully controlled
 in the counting facilities. Intense $^{51}$Cr sources \cite{Cr51} have been used to calibrate the sensitivity of the
 detector to neutrinos.

The results from these experiments to date \cite{SAGE,GNO} are as follows:
GALLEX + GNO: $70.8 \pm 4.5 (statistical) \pm 3.8 (systematic)$ SNU, SAGE: $70.9 +5.3/-5.2 (stat) +3.7/-3.2 (syst)$ SNU. These numbers are in excellent agreement and are much smaller than the predictions of the standard solar model (131 SNU)~\cite{BPG04}.

\subsection{Real-Time Solar Neutrino Detectors}

\subsubsection{Kamiokande and Super-Kamiokande}

From 1983 to 1996 the 1 kiloton (680 ton fiducial) light water neutrino detector, Kamiokande~\cite{KAMIOKA}
 (1000 meters underground, 2600 m.w.e.)
in the Kamioka mine in Japan, studied solar neutrinos, as well as other physics processes such as proton decay and
atmospheric neutrinos. In 1996, the much larger (22.5 kton light water) Super-Kamiokande detector~\cite{SKNIM}
 began operation at the same site, studying solar neutrinos and other physics with higher sensitivity.
Solar neutrinos are observed through the elastic scattering
 of neutrinos from electrons, observed by the detection of Cerenkov light from the recoiling electron. The energy threshold for Kamiokande was about 7 MeV and about 5 MeV for Super-Kamiokande. Therefore the sensitivity of these detectors is almost entirely for $^8$B solar neutrinos with a tiny contribution from the $^3$Hep reaction.

For Kamiokande, the flux of neutrinos from the $^8$B decay was measured~\cite{KAMIOKA} to be
 $2.8 \pm 0.19 (stat) \pm 0.33 (syst) \times 10^{+6}$ cm$^{-2}$ s$^{-1}$, clearly less than the solar model
 prediction~\cite{BPG04} of $5.82 (1 \pm 0.23) \times 10^{+6}$ cm$^{-2}$ s$^{-1}$. Studies of the variation in flux as a function of time showed no statistically significant effects.

\begin{figure}
\begin{center}
\includegraphics[width=4 in]{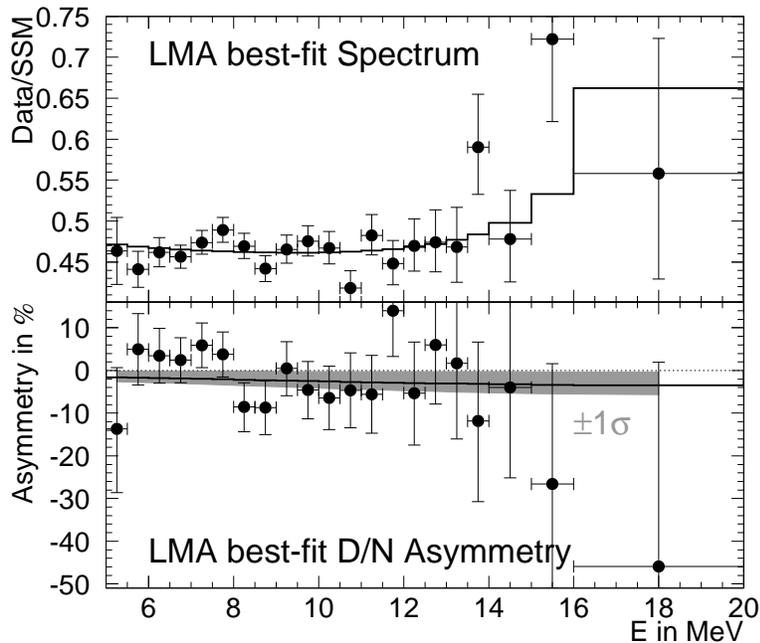}
\caption{\label{fig:skspectrum}
Solar neutrino elastic scattering energy spectrum (top) and D/N Asymmetry (bottom) as measured by Super-Kamiokande.
 The predictions (solid lines) are for the LMA solution. The gray bands are the $\pm \sigma$ ranges for the
asymmetry over the entire energy range.}
\end{center}
\end{figure}

 The solar neutrino flux from $^8$B decay measured by Super-Kamiokande~\cite{SK} is
 $2.35 \pm 0.02 (stat) \pm 0.08 (syst) \times 10^{+6}$ cm$^{-2}$ s$^{-1}$, and is also significantly
 smaller than the solar model prediction. The neutrino energy spectrum is very similar to that expected
 from $^8$B decay and no distortion due to the MSW effect~\cite{MSW} has been observed. Figure~\ref{fig:skspectrum} shows a comparison of the observed spectrum with the spectrum from $^8$B decay.

To look for indications of the MSW effect through regeneration of neutrinos while passing through the Earth~\cite{DayNight}, an analysis
of the spectra including solar zenith angle has been performed. Figure~\ref{fig:skspectrum} also shows the spectra for day and night and it is clear that there is no significant difference observed.
A maximum likelihood analysis of this 1496 days of data shows a day-night asymmetry of
$ -1.8 \pm 1.6 (stat.) ^{+1.3} _{-1.2} (syst.) \%$, consistent with zero and with the expected value (2.1) 
from the most recent analysis of the LMA solution for neutrino oscillations as discussed in reference \cite{BPG04}.
A further analysis of the data was carried out, seeking distortion of the $^8$B energy spectrum from magnetic scattering associated with a finite neutrino magnetic moment. Potential distortions of the weak ES spectrum from oscillation of $^8$B neutrinos were taken into account by incorporating data from other solar neutrino experiments and an upper limit for the neutrino magnetic moment of 
$1.1 \times 10^{-10} \mu_B$ was obtained for 90 \% confidence level~\cite{SKMagMom}.

\begin{figure}
\begin{center}
\includegraphics[width=4 in]{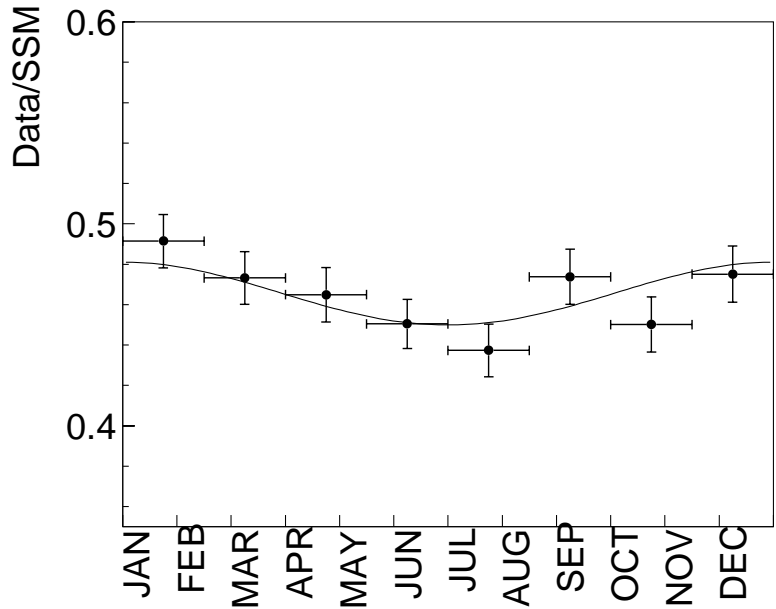}
\caption{\label{fig:skseasonal}
Seasonal variation of the solar neutrino flux as measured by Super-Kamiokande. The curve shows the expected seasonal variation of the flux introduced
by the eccentricity of the Earth's orbit.}
\end{center}
\end{figure}

If there were a significant neutrino magnetic moment, this could couple with solar magnetic field distributions
 and produce different observed fluxes through flavor change as neutrinos pass through different solar regions at
 various times of year. However, the neutrino flux as a function of time of year is observed by the Super-Kamiokande measurements to follow the variation
 expected for the Earth-Sun distance variation with no observable additional effects as shown in Figure~\ref{fig:skseasonal}.
With the full 1496 day data set, the amplitude of the neutrino flux variation is measured to be $1.51 \pm 0.43$ (consistent with one) times the amplitude expected from the eccentricity of the Earth's orbit. The perihelion shift is measured to be $13 \pm 17$ days (consistent with zero). 
Studies over longer periods of time to seek correlations between the data and the sunspot numbers (indicative of magnetic fields in the outer regions of the Sun) also show no correlations~\cite{SK}.

\subsubsection{Sudbury Neutrino Observatory}

The principal objective of the Sudbury Neutrino Observatory (SNO) has been the clear observation of flavor change for $^8$B solar
neutrinos by a comparison of the total flux of active neutrinos from decay with the total flux of electron neutrinos.
The measurement of flavor change does not rely on solar model calculations and in fact,
the total flux of active neutrinos can be compared with the predicted flux of $^8$B neutrinos as a test of those
calculations. The fluxes are measured through the following reactions in a heavy-water-based Cerenkov detector:

 \begin{center}
  \begin{tabular}{ll}
     $\nu_e + d \rightarrow p + p + e^-$\hspace{0.5in} & (Charged Current:CC)\\
     $ \nu_x + d \rightarrow p + n + \nu_x$ & (Neutral Current:NC)\\
     $ \nu_x + e^- \rightarrow \nu_x + e^-$  & (Elastic Scattering:ES)\\        
  \end{tabular}
 \end{center}

The CC reaction is sensitive exclusively to electron neutrinos,
 while the NC reaction is sensitive to all neutrino flavors ($x = e, \mu, \tau$) above the energy threshold of
 2.2 MeV. The ES
 reaction is sensitive to all flavors as well, but with reduced sensitivity to $\nu_{\mu}$ and $\nu_{\tau}$.
 Comparison of the $^8$B flux deduced
 from the CC  reaction with the total flux of active neutrinos observed with the NC reaction (or the ES reaction with
 reduced sensitivity) can provide clear evidence of neutrino flavor transformation without reference to solar model
 calculations.

The SNO detector~\cite{SNONIM} uses 1100 tons of heavy water
 sited 2000 meters underground (6200 meters water equivalent) near Sudbury, Ontario, Canada. Neutrinos are detected
 via the Cherenkov process in the central heavy water volume or in a surrounding volume
 of light water. The CC and ES reactions are observed through the Cherenkov light produced by the electrons. The NC reaction is observed
 through the detection of the neutron in the final state of the reaction.  The SNO experimental plan involves
 three phases
 wherein different techniques are employed for the detection of neutrons from the NC reaction. During the first phase,
 with pure heavy water, neutrons were observed through the Cherenkov light produced when neutrons are captured on
 deuterium, producing 6.25 MeV gammas. For the second phase, about 2.7 tons of salt was added to the heavy water
 and neutron detection was enhanced through capture on Cl, with about 8.6 MeV gamma energy release and higher
 capture efficiency. The neutrons from the NC reaction capture on Cl
and provide a more isotropic pattern at the PMT's for these events than for the CC events. This enables the fluxes
 from the CC and NC reactions to be extracted on a statistical basis with no constraint on the shape of
the CC energy spectrum. For the third phase, the salt was removed and an array of $^{3}$He-filled proportional counters
has been installed to provide neutron detection independent of the PMT array.

During operation to date, radioactivity
 levels were achieved that result in background events for the NC reaction from photodisintegration of deuterium by gamma
 rays that are less than 5\% of the predicted rate from the Standard Solar Model neutrino flux. 
The threshold for observation of the CC and ES reactions is 5.5 MeV in equivalent electron energy. 

\begin{figure}
\begin{center}
\includegraphics[width=3in]{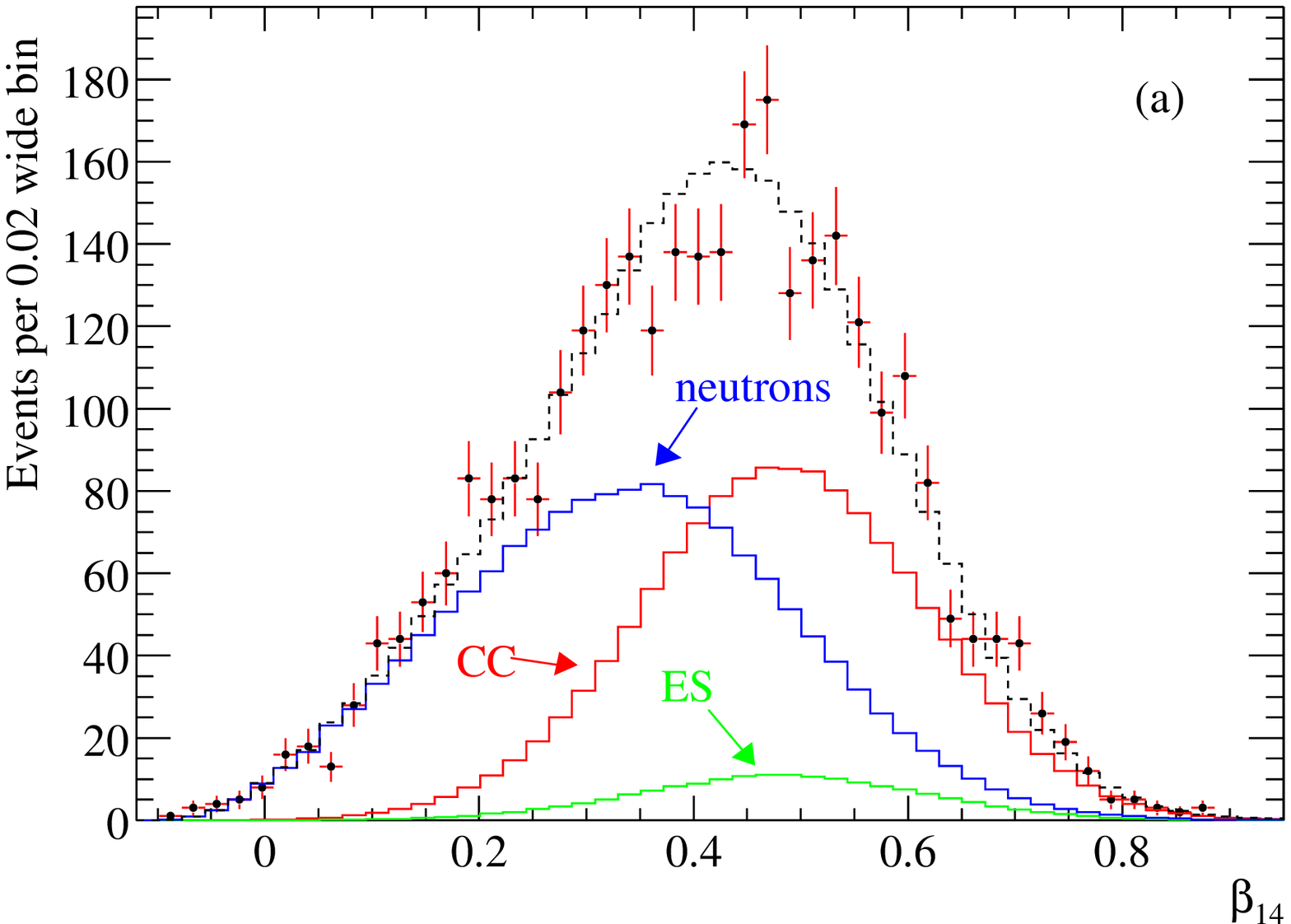}
\includegraphics[width=3in]{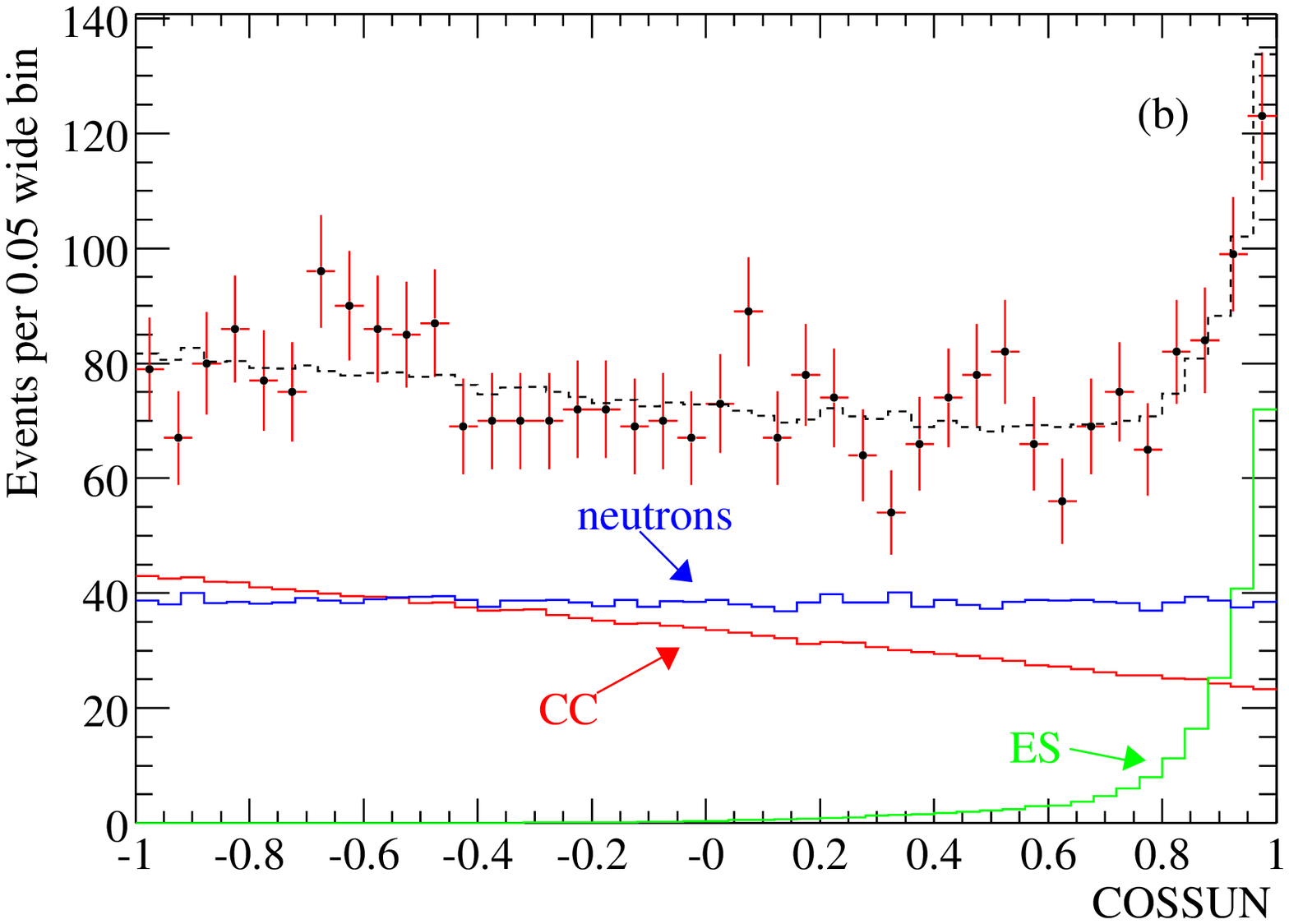}
\includegraphics[width=3in]{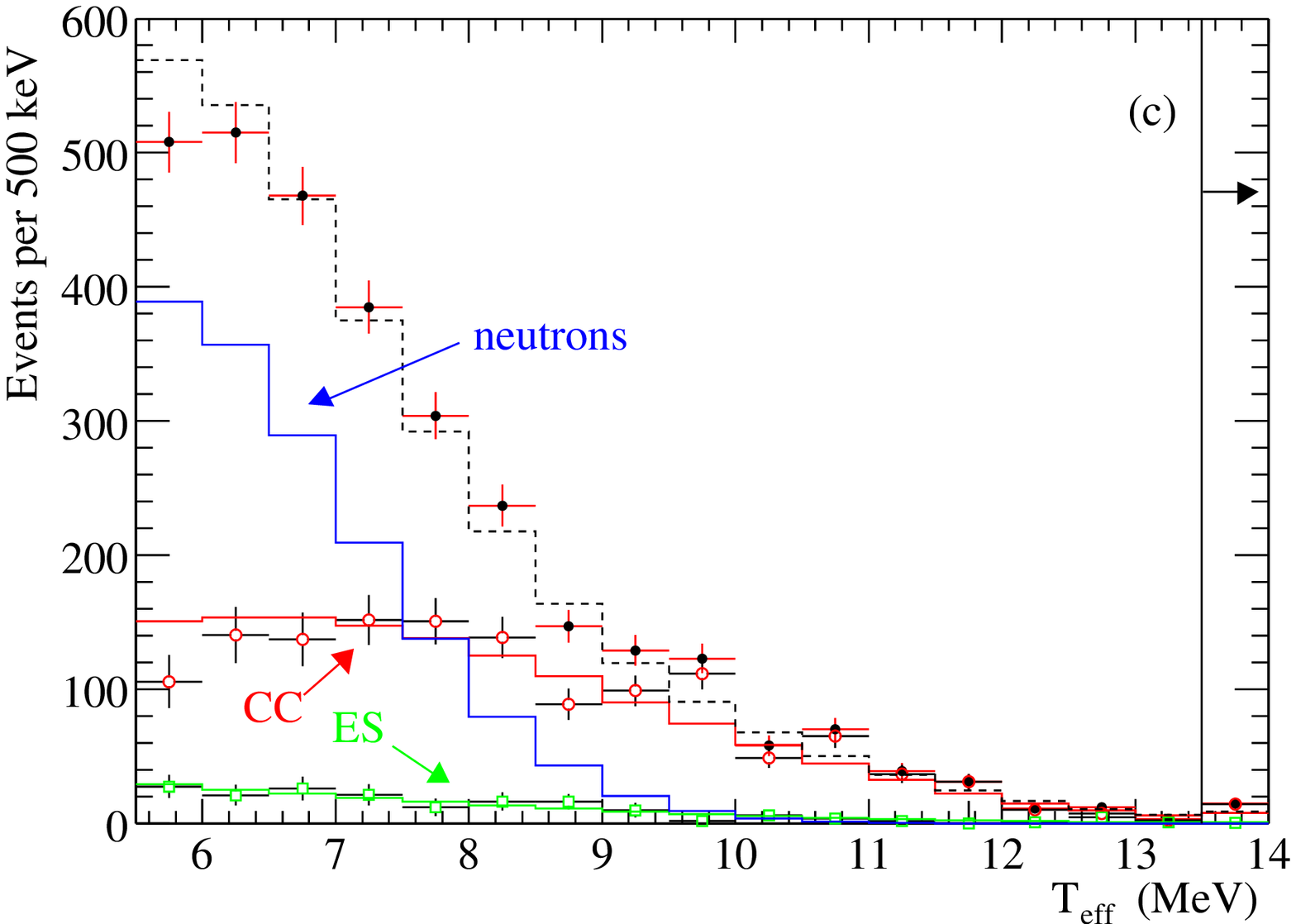}

\caption{\label{fig:salt_data}
Distribution of (a) $\beta_{14}$, (b) $\cos \theta_{\odot}$ and
 (c) kinetic energy, for the selected events.  The CC and ES spectra are extracted from the data
 using $\beta_{14}$ and $\cos \theta_{\odot}$ distributions in each energy bin.   Also shown are the
 Monte Carlo predictions for CC, ES, NC + internal and external-source neutron events, all scaled to the fit
 results.  The dashed lines represent the summed components.  All distributions are for events with
 $T_{\rm eff}$$\geq$5.5 MeV and R$_{\rm fit}$$\leq$ 550 cm. Differential systematics are not shown.}
\end{center}
\end{figure}

Results from the initial phase of the experiment with pure heavy water~\cite{SNO1,SNO,SNO3} showed
 clear evidence for neutrino flavor change, without reference to solar model calculations. The hypothesis of no
 flavor change was tested and found to be violated using summed data for the CC and NC reactions observed through Cerenkov processes
in the heavy water. With the addition of salt it was possible to break the strong correlation between the CC and NC
 event types because the pattern of light on the PMT's is different for the two reactions.
An event angular distribution parameter $\beta_{14}$ (defined in reference~\cite{SNOsalt}) was used to
separate NC and CC events on a statistical basis. The data from the salt phase is shown in Figure~\ref{fig:salt_data},
together with the best fit to the data of the NC (6.25 MeV gamma) shape, the CC and ES reactions 
and a small background component determined from independent measurements.

The fluxes inferred for this fit are:
\begin{eqnarray*}
\phi^{SNO}_{CC} & = & \snoccfluxunc \\
\phi^{SNO}_{ES} & = & \snoesfluxunc \\
\phi^{SNO}_{NC} & = & \snoncfluxunc. 
\end{eqnarray*}
\noindent
where these fluxes and those following in this section are quoted in units of 10$^{6}$ cm$^{-2}$ sec$^{-1}$.

By comparison of the CC and NC fluxes it is clear that only about 1/3 of the neutrinos reaching the Earth are electron
neutrinos. These results violate the hypothesis test for no flavor change at greater than 7 $\sigma$. Previous measurements with pure heavy water had reached similar conclusions of flavor change at levels of 5.5 $\sigma$~\cite{SNO, SNO3} (by consideration of the summed CC and NC and ES data) and 
3.3 $\sigma$~\cite{SNO1} (by comparison of the CC data with ES measurements by Super-Kamiokande). The measurements from the salt phase, combined with the results of other solar neutrino experiments indicate that the mixing is non-maximal (mixing angle less than $\pi$/4) with a confidence level corresponding to 5.5 standard deviations. The measured total flux of active ${}^{8}B$ neutrinos measured with the NC reaction in the salt phase is in very good agreement with and more accurate than the SSM prediction~\cite{BPG04} of $5.82 (1 \pm 0.23)$.

\subsection{Summary of Results for Solar neutrino measurements to date}
\begin{figure}
\begin{center}
\includegraphics[width=3.73in]{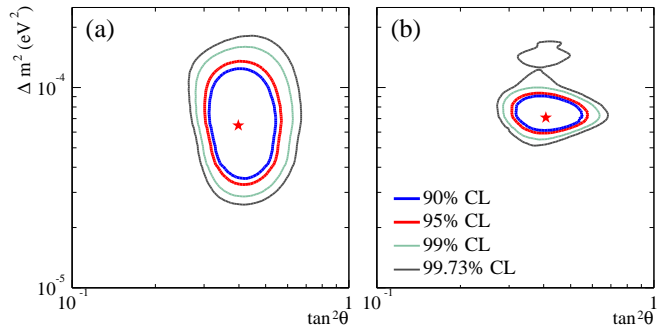}
\caption{\label{globalmsw}Global neutrino oscillation contours. (a) Solar global: D$_2$O day and night spectra, salt CC, NC, ES fluxes, SK, Cl, Ga.  The best-fit point is $\Delta m^2=6.5\times10^{-5}$, $\tan^{2}\theta=0.40$, $f_{B}=1.04$, with $\chi^{2}$/d.o.f.=70.2/81. (b) Solar global + KamLAND.  The best-fit point is $\Delta m^2=7.1\times10^{-5}$, $\tan^{2}\theta=0.41$, $f_{B} = 1.02$.  In both (a) and (b) the ${}^{8}$B flux is free, the {\em{hep}} flux is fixed at its solar model value and the other solar fluxes are included with the central value and uncertainty calculated with the solar model. }
\end{center}
\end{figure}

The measurements to date have provided convincing evidence for flavor change of neutrinos and matter enhancement
 of these effects in the Sun. The data to date for solar neutrinos has been analyzed by many authors. Figure~\ref{globalmsw}(a) shows the results for an analysis allowing for two neutrino types~\cite{SNOsalt}, following the SNO salt data.

Measurements of reactor anti-neutrinos by the KamLAND detector reported in 2002 show oscillation parameters matching well with the LMA region for the solar neutrino case (assuming CPT conservation). Figure~\ref{globalmsw}(b) shows the remaining acceptable region when the KamLAND data is combined with the solar data.

The KamLAND detector~\cite{KAMLAND} is a 1 kton liquid scintillator detector constructed in the former Kamiokande location with the initial objective to look for oscillations for neutrinos produced in the power reactors in Japan and Korea. Work is in progress to reduce the radioactive background in the detector and use it for measurements of low energy solar neutrinos. These measurements will be discussed below.

The observation of anti-neutrino disappearance with similar oscillation parameters to those observed for solar neutrinos places strong restrictions on flavor change mechanisms other than oscillation arising from finite neutrino mass. Mechanisms such as resonant spin-flavor procession, flavor-changing neutral currents and neutrino decay are more strongly disfavored as explanations for the low fluxes observed on Earth for solar electron neutrinos. Oscillation to sterile neutrinos is also strongly disfavored and sub-dominant oscillations to sterile neutrinos are restricted by comparison of the active total flux of $^8$B solar neutrinos observed by the NC reaction in SNO with the $^8$B flux calculated with solar models. If CPT conservation is assumed, a limit on sub-dominant sterile components can be obtained without reference to solar model calculations~\cite{Bahcallsterile} by using KamLAND data combined with SNO CC and NC measurements. Conceptually, the oscillation parameters defined by Kamland are used to determine the initial $^8$B solar neutrino flux from the SNO CC measurement. That flux can then be compared with the NC measurement to set limits on oscillation to sub-dominant sterile components. 

The KamLAND data has also been analysed~\cite{KamAntinu} to place restrictions on electron anti-neutrinos from the Sun. This provides strong restrictions on processes that change $^8$B solar electron neutrinos into anti-neutrinos, including spin-flavor precession and neutrino decay. The analysis sets a limit of $2.8 \times 10^{-4}$ of the Standard Solar Model flux for the flux of electron anti-neutrinos (90 \% C. L.)

\section{Future solar neutrino experiments}
\subsection{Physics Motivation}
As can be seen from Figure~\ref{globalmsw}, the solar measurements to date, combined with KamLAND are providing strong constraints on the parameters for the mixing of active neutrinos. This is discussed in more detail in the following paper~\cite{BPG04}. The main physics motivations for future solar neutrino measurements are as follows:
\begin{enumerate}

	\item {\it Improvement in accuracy for the mixing parameters for three active neutrino types}

The improvement of the 1-2 mixing parameters provides an important objective for future measurements of solar neutrinos. In addition, improvement in the accuracy for the solar neutrino measurements leads to improvement in the knowledge of other parameters in the full MNSP mixing matrix~\cite{MNSP} for three active neutrinos. For example, as discussed by Maltoni et al.~\cite{Maltoni} measurements of solar neutrinos and KamLAND provide the dominant constraints on the $\theta_{13}$ parameter for the smallest part of the allowed range of $\Delta m^2_{23}$ for fits to the atmospheric neutrino measurements. Therefore improvements in solar neutrino measurements can provide valuable information prior to direct measurements of the $\theta_{13}$ parameter that are planned for the future. 

\item {\it Further study of matter-enhanced oscillation}

As shown in Figure 1 of the following paper by Bahcall and Pena-Garay~\cite{BPG04}, the oscillation situation for the LMA parameters is very different for neutrino energies above a few MeV where matter enhanced oscillation dominates and below this energy, where the oscillation is close to the vacuum case. Therefore, it is of interest to study the energy dependence for solar neutrinos to confirm this present understanding of the oscillation situation for LMA parameters as well as to provide more accurate parameters for the oscillation. 

A charged current and a neutral current measurement for the same neutrino reaction below a few MeV would provide a measure of electron neutrino survival without reference to solar model calculations. This survival probability could then be compared with the SNO results for $^8$B neutrinos to demonstrate directly the difference for vacuum oscillation and the energy-dependent matter enhancement. Improvement in the accuracy of existing experiments could provide direct evidence for matter enhancement in the Sun through the distortion of the neutrino spectrum for $^8$B neutrinos or a clear day-night difference in observed flux, arising from matter-enhancement in the Earth. The effects are expected to be small for the LMA solution so the measurements are difficult.

\item {\it Search for physics beyond three active neutrino types}

In addition, improvements in the 1-2 parameters are of value in seeking evidence for physics beyond three active neutrino types, through tests of the unitarity of the MNSP matrix. At present, these tests are limited by the lack of knowledge of $\theta_{13}$ but when future measurements of that parameter are made, the unitarity of the MNSP matrix will place restrictions on possible contributions from sterile neutrinos and other physics, such as CPT violation when KamLAND data is included~\cite{CPT, PakvasaBarger}. In addition, as has been pointed out by some authors~\cite{SmirnovSterile}, distortion of the energy spectrum could be possible for certain cases involving sterile neutrinos.

\item {\it Further Information on Solar Models}

With a clear knowledge of the oscillation parameters for solar neutrinos, further information could be obtained on solar models by individual observations of pp, $^7$Be, pep and CNO neutrinos. Although solar models appear to work very well, improved accuracy on neutrinos from these reactions can test model parameters with extreme accuracy, including radial profile information, since the varioous reactions are concentrated in different radial sections of the Sun.

\end{enumerate}

This physics motivation leads to the following experimental objectives for future measurements of solar neutrinos:
\begin{enumerate}
	\item Improvement in the accuracy of currently running solar neutrino experiments to reduce the experimental uncertainties and improve the restrictions on the 1-2 mixing parameters. (This also is the objective of the KamLAND experiment with reactor anti-neutrinos.)
	\item Measurements of lower energy solar neutrinos (pp, pep, $^{7}$Be, CNO) including real-time measurements via CC and ES reactions. The objective would be to perform a measurement of neutrino flavor change for neutrinos with energy below a few MeV without reference to solar model calculations. These measurements of solar electron neutrino fluxes for the individual low-energy reactions will also test solar models in much greater detail. 
	\item Study of the ES process for the lowest energy solar neutrinos to measure the neutrino magnetic moment with an accuracy greater than present measurements with terrestrial or solar neutrinos.
\end{enumerate}

\subsection{General Description of Experiments}

\subsubsection{Target Materials}

In addition to the materials in present use, (water, heavy water, gallium, formerly chlorine), there are a number of materials proposed for future experiments whose properties are matched to the objectives for lower energy neutrinos. For the charged current inverse electron-capture 
reaction $\nu_e + (A,Z) \rightarrow e^- + (A,Z+1)^* $ the nuclear properties are very important and targets with low energy thresholds for inverse beta decays of electron neutrinos are being considered, including $^{115}$In, $^{100}$Mo, $^{82}$Se, $^{160}$Gd, $^{176}$Yb and $^{7}$Li. For the ES reaction, $\nu_{e,\mu,\tau} + e^- \rightarrow \nu_{e,\mu,\tau}$ 
hydrocarbon liquid scintillators are considered in order to obtain significantly more light than can be obtained from the Cerenkov process. 

However, for pep and CNO neutrinos, background from cosmogenically produced $^{11}$C can be a problem requiring an muon veto and imposition of dead time following an event, except perhaps for sites as deep as SNO and its future extension, SNOLAB. For pp neutrinos, naturally occurring $^{14}$C is prohibitive in hydrocarbons and so several proposed experiments are based on other materials such as liquid He, Ne or Xe or gaseous He. Very accurate measurements are necessary to measure the total flux of all active neutrinos, because the Neutral Current interaction only contributes about $1/6$ of the ES cross section.

For ES measurements, the expected interaction rates corresponding to about 0.6 of the SSM are about 1.2/ton/day for pp-neutrinos, 0.024/ton/day for pep-neutrinos and 0.3/ton/day for $^7$Be-neutrinos. Therefore, detector masses of 20 to 500 tons are required to obtain event rates greater than about 10 per day for these solar reactions. The CC reaction rates depend on the matrix elements for the inverse electron capture processes. As an example, the interaction rate in $^{115}$In for 0.6 of the SSM would be about 0.2/ton/day for pp-neutrinos and about 0.04/ton/day for $^7$Be-neutrinos.

\subsubsection{Radioactive Backgrounds}

The neutrino signals of interest, below a few MeV, are inevitably affected by radioactive backgrounds from naturally occurring radioisotopes in detector materials ($^{14}$C, $^{40}$K, $^{238}$U, $^{232}$Th and related decay products), from cosmogenically produced isotopes ($^{11}$C and $^{39}$Ar) and other man-made isotopes from the atmosphere ($^{85}$Kr). Control of these backgrounds is accomplished by extreme care in detector manufacture, as well as the use of very deep locations and in-situ purification techniques.

For ES measurements, the signal is a single recoil electron. Therefore, background from radioactive decays can mimic a neutrino signal directly. As an example, for a signal to background of about 1 for $^7$Be-neutrinos, Th and U impurity levels of less than about $5 \times 10^{-17}$g~U(Th)/g(hydrocarbon) are required. For CC measurements, there are coincidence requirements for the signals that can result in somewhat less stringent radioactive impurity requirements.

\subsection{Constructed Experiments}

SAGE, Super-Kamiokande and SNO are continuing operation and GNO will finish its measurements in the near future. In addition, KamLAND is working on the reduction of radioactive impurities with the objective of performing measurements of $^7$Be-neutrinos. The BOREXINO experiment~\cite{BXAP16,BXAP18} is a 300-ton hydrocarbon liquid scintillation detector that has been constructed in the Gran Sasso laboratory with the primary objective of measuring $^7$Be-neutrinos. Extensive work has been done on the reduction of radioactive impurities as demonstrated in preliminary measurements in a smaller test detector~\cite{ctf}. It is expected that detector filling for Borexino could take place later in 2004.

The SAGE detector will be continuing with sensitivity in the pp-neutrino region. Improved accuracy for this region will be valuable for further restricting oscillation parameters. The Super-Kamiokande detector will shut down in 2005 to refurbish the phototubes destroyed in the accident in 2002. This will re-establish the phototube coverage that was being used prior to the accident and should restore the analysis threshold for solar neutrinos to 5 MeV.

The SNO detector is entering the third phase, using $^3$He-filled proportional counters to make a fully independent measurement of the neutrons from the NC reaction. About 2.5 years of operation in this configuration is planned. Projected accuracy for the NC measurement is expected to be improved by about a factor of two and the removal of the correlated signals should also improve the accuracy of CC flux measurement by about 1.5 and provide more sensitivity for the study of $^8$B spectral distortion. With this improvement in the SNO results and with the future running planned for the KamLAND experiment and other solar neutrino experiments, one can expect that the allowed region in Figure~\ref{globalmsw}(b) will be reduced on both axes, with the improvements in SNO reducing the $\theta_{12}$ range by about a factor of two and improvements in KamLAND accuracy from future running with reactor neutrinos reducing the range of $\Delta m_{12}^2$ by about a factor of three~\cite{Yuksel} in about three years time. This improvement in the accuracy of the 1-2 parameters will provide significant constraints on $\theta_{13}$ and on sterile neutrinos as discussed above. In addition, further measurements of low energy solar neutrinos by experiments to be discussed below all improve on this accuracy, particularly if it is possible to measure both CC and NC processes for these low energy cases.

The KamLAND detector is improving the purification of their 1000 ton central volume of liquid scintillator with the particular objective of measuring $^7$Be neutrinos~\cite{Furuno}. The principal backgrounds that must be reduced are $^{210}$Pb, $^{85}$Kr and $^{40}$K that affect the region below 1 MeV, as well as U and Th chain elements, particularly Rn and $^{210}$Pb. Measurements of U and Th chain elements in the fiducial volume indicate impurity levels at present of about $5 \times 10^{-17}$ g(Th)/g and $4\times 10^{-18}$ g(U)/g (equilibrium equivalents). Techniques for reduction of these contaminants from the liquid scintillator are being developed, using gas stripping with nitrogen gas that has been purified of Kr, Ar and Rn, as well as stripping of $^{210}$Pb with extraction techniques with ultra-pure water. The expected fiducial volume for solar neutrinos is on the order of 500 tons. The relatively shallow depth of the KamLAND site will probably prevent observation of pep and CNO solar neutrinos due to $^{11}$C background induced by muons.

The Borexino detector~\cite{BXAP16} should be in operation in 2004, incorporating the many radiopurification techniques developed for this experiment that has always had a primary aim of measuring the flux of $^7$Be neutrinos. The Borexino collaboration have pioneered many techniques for reducing radioactive background in liquid scintillator and used a Counting Test Facility~\cite{ctf} to obtain the design objectives for a number of the radioactive contaminants. The results from the CTF have been used to develop purification techniques (including distillation of the scintillator as well as water and nitrogen stripping) that will be applied for the full scale detector. Techniques~\cite{Schoenert} for isolating production locations for muon-induced $^{11}$C may make it possible to observe pep and CNO neutrinos if other radioactivity, particularly $^{210}$Pb can be reduced sufficiently. Radiopurity objectives are $ 10^{-16}$g/g for uranium and thorium, and their progenies (assuming secular equilibrium), $ 10^{-14}$g/g for potassium, $ 10^{-10}$g~/g for argon and $ 4\times 10^{-16}$g/g for krypton to obtain background rates less than the expected signals for $^7$Be.

The possibility of replacing the heavy water in the SNO detector with liquid scintillator is being considered as an option when the heavy water phase is completed. The deep location would provide a significant reduction of $^{11}$C for potential measurements of pep and CNO solar neutrinos and there could be other advantages for observation of geoneutrinos, reactor neutrinos and possibly double beta decay with added isotopes. 

\subsection{Next generation experiments}

Future projects under development are aimed primarily at measurements of pp and $^7$Be neutrinos by the CC and ES reactions. Table~\ref{tab:projects} lists the largest of the projects under development.

\begin{table*}
\caption{\label{tab:projects}
	Future Projects.}
\begin{tabular}{lll}
	\hline Experiment & Reaction & Target and Technique \\
	\hline XMASS~\cite{xmass} & ES  & Liquid Scintillator (Xe) \\
	HERON~\cite{heron} & ES  & Liquid Scintillator (He)   \\ 
	CLEAN~\cite{clean} & ES  & Liquid Scintillator (Ne)\\
	TPC~\cite{tpc} & ES  & Gas Time Projection Chamber(He + Hydrocarbon) \\
	LENS~\cite{lens,RaghavanNoon} & CC: $^{115}$In,$^{176}$Lu & Liquid Scintillator (Metallo-organic) \\
	MOON~\cite{moon} & CC: $^{100}$Mo & Fibre-optic and/or Liquid Scintillator (Metallo-organic)\\
	\hline
\end{tabular}
\end{table*}

{\it ES Measurements}

Measurements at the Borexino CTF~\cite{C14} indicate $^{14}$C/$^{12}$C ratios of about 10$^{-18}$
in organic scintillators. This background is a problem for the measurement of pp-neutrinos in ES experiments 
because of the 156 keV $\beta$-decay endpoint for $^{14}$C. Therefore a number of experiments under development 
are based on inorganic materials, principally a noble gas scintillator. Such scintillators can provide over 4 times as many photons as a typical liquid scintillator but the photons are at UV wavelengths and so wavelength shifters are required.

The XMASS experiment~\cite{Namba} proposes to use 10 tons of liquid Xenon to observe pp and $^7$Be solar neutrinos via scintillation light. This mass could provide 10 events/day for pp and 5 events per day for $^7$Be neutrinos. The experiment is being developed in stages. Presently a 100 kg prototype detector is in operation and has confirmed initial Monte Carlo calculations for detector performance. It is proposed to build a 1 Ton detector before moving to the full scale experiment. The detector is located in an ultra-clean facility in the Kamioka mine near the Super-Kamiokande detector. For the 10 ton detector a 2.5 meter diameter spherical design is planned using low-background steel photomultiplier tubes with quartz windows that can operate at liquid xenon temperature. Distillation techniques have been developed and applied to reduce $^{85}$Kr background by up to a factor of 300 in one pass. The two-neutrino double beta decay of $^{136}$Xe has not be measured to date and will be one of the objectives of the first two phases. If its magnitude is too high, isotope separation may be necessary to remove the $^{136}$Xe for the solar neutrino measurements. For 5 years of operation of the full scale XMASS detector, for the detection of pp-neutrinos, Nakahata has indicated~\cite{Nakahata} an expected sensitivity to $sin^2 \theta_{12}$ of about $\pm 0.03$.

The HERON experiment~\cite{heron} design uses liquid Helium with detection via scintillation and/or observation with wafer calorimeters of rotons or 'electron-bubbles' formed at the surface of the helium. Tests have been performed of this innovative technology and detector simulations performed including background estimates. The use of helium provides an inherently low-background medium for the measurements. A 20-ton detector is planned with pp-neutrino rates of more than 1000/year in the fiducial volume. With spectral measurements on the pp spectrum, the shape could provide a sensitive probe of neutrino magnetic moment.

Liquid neon is the basis for the CLEAN experimental design~\cite{clean}. Neon has no long-lived radioactive isotopes, is an excellent scintillator, has a higher density than liquid helium, can be readily purified and is relatively inexpensive. A spherical design with a full radius of 3 meters and a fiducial radius of 1.5 meters is being considered. Monte Carlo simulations indicate that the detector would be capable of sensitive dark matter measurements as well as accurate pp- and $^7$Be flux measurements.

A large (18 m $\times$ 20 m) high-pressure (10 Atmosphere) Time Projection Chamber using 97 \% helium gas has been proposed~\cite{tpc} to measure the spectra for low-energy solar neutrinos. The detector technology would build on the MuNu detector experience and could provide directional identification of the ES reaction through its strong angular distribution for the recoil electrons. As with the other detectors, dark matter detection would also be one of the objectives of the design.

{\it CC Measurements}

The LENS experiment~\cite{lens} has been working with a low-background test facility to develop a CC measurement based on $^{115}$In (with a threshold energy 120 keV) in a metallo-organic scintillator. Scintillators have been developed with over 5 \% loading by weight of In and acceptable light attenuation lengths. A detector with high granularity is required to identify the background events. Therefore the detector design has many modules. The neutrino events will be tagged by two gammas from a cascade in the final nucleus, reducing background from radioactivity significantly. A total volume of about 60 tons of Indium in 3000 tons of liquid scintillator is planned~\cite{RaghavanNoon}. This would provide about 3 events per day for either a pp or a $^7$Be flux at 0.6 of the Standard Solar Model. The nuclear matrix element for the low energy neutrino absorption reaction has been measured through the (p,n) reaction. However, to provide an accurate calibration, an measurement with an $^{37}$Ar radioactive neutrino source is planned.

The MOON~\cite{moon} experiment proposes to use a low energy inverse electron capture reaction on $^{100}$Mo to observe pp neutrinos with excellent sensitivity and a threshold energy of 168 keV. The daughter nucleus $^{100}$Tc beta decays with a half life of 16 seconds, so the signature of the CC reaction is two localized electrons. $^{100}$Mo itself undergoes two-neutrino double beta decay at a rate that constitutes a background for the detection of solar neutrinos and so a detector with small modules is used to discriminate double beta decays from neutrino events. A design is under development using optical fibers for an extensive modular structure. A counting rate of 0.4 events per day for pp-neutrinos is projected for a detector containing about 3 tons of $^{100}$Mo isotope.

\subsection{Summary}

Measurements of solar neutrinos have contributed substantially to our knowledge of neutrino properties and of the physics of the Sun. Neutrino flavor change has been clearly observed in an appearance measurement. A series of sensitive measurements with different energy thresholds have tested the Standard Solar Model for the dominant reactions and there is very good agreement. The measurements to date, combined with reactor anti-neutrino measurements have defined the flavor change mechanism to be the oscillation of massive neutrinos and defined the acceptable range of oscillation parameters with reasonable accuracy. They have also provided limits at the 10 \% level for sub-dominant components such as sterile neutrinos. 

Future experiments will provide further accuracy for neutrino oscillation parameters as well as providing the first measurements of individual neutrino reactions at low energy to probe solar models with greater sensitivity.


\section*{References}

\begin{thebibliography}{999}

\bibitem{BPG04} Bahcall, J.N and Pena-Garay, C., this volume.

\bibitem{davis1} Davis et al., {\it Phys. Rev. Lett.} {\bf 14} (1968) 20;
 B.T. Cleveland, T. Daily, R. Davis, Jr., J.R. Distel, K. Lande,
 C.K. Lee, P.S. Wildenhain, J. Ullman, {\it Astrophys. J.}{\bf496} (1998) 505. 

\bibitem{SAGE} V.N. Gavrin et al. {\it Nucl. Phys. B (Proc. Suppl.)}{\bf 118} (2003) 39.

\bibitem{GNO} W. Hampel et al. {\it Phys. Lett.} {\bf B 447} (1999) 127 ; T. Kirsten {\it Proceedings of the Neutrino 2002 Conference, Munich} (2002)

\bibitem{KAMIOKA} Y. Fukuda et al. {\it Phys. Rev.}{\bf D 44} (1996) 1683.


\bibitem{SK} S. Fukuda et al. {\it Phys. Rev. Lett.}{\bf 86} (2001) 5651; S. Fukuda {\it et al.}, {\it 
Phys. Lett. B} {\bf 539} (2002) 179; M.Smy, for the Super-Kamiokande collaboration,
{\it Nucl. Phys. B (Proc. Suppl.)} 118 (2003) 25; M. B. Smy et al {\it Phys. Rev. Lett.} {\bf XX} (2003) XXX.      

\bibitem{SNOsalt} S. N. Ahmed et al. , nucl-ex/0309004.

\bibitem{MSW}S.~P. Mikheyev and A.~Y. Smirnov, in
  Massive Neutrinos in Astrophysics and in
  Particle Physics, Proceedings of the Moriond Workshop, edited by
  O. Fackler and J. Tran Thanh Van, Editions Fronti\`eres,
  Gif-sur-Yvette, 335 (1986); S.P.Mikheyev and A.Yu. Smirnov, {\it Sov. J. Nucl. Phys.} {\bf 42} (1985) 1441;
L. Wolfenstein, {\it Phys. Rev.} {\bf D 17} (1978) 2369. 

\bibitem{NuDecay} J. F. Beacom and N.F. Bell {\it Phys. Rev.} {\bf D 65} (2002) 113009.

\bibitem{RSFP} O. G. Miranda et al. {\it Nucl. Phys.} {\bf B 595} (2001) 360; B.C. Chauhan and J. Pulido, {\it Phys. Lett.} {\bf 66} (2002) 053006; E. Kh. Akhmedov and J. Pulido, {\it Phys. Lett.} {\bf B 553} (2003) 7.

\bibitem{helio} J. N. Bahcall et al. {\it Phys. Rev. Lett.} {\bf 78}, (1997) 171. 

\bibitem{MNSP} Z. Maki, N. Nakagawa and S. Sakata, {\it Prog. Theor. Phys}
{\bf 28} (1962) 870, B. Pontecorvo {\it J. Expt. Theoret. Phys}{\bf 33} (1957) 549
{\it J. Expt. Theoret. Phys}{\bf 34} (1958) 247, V. Gribov and B. Pontecorvo {\it Phys. Lett.}{\bf B28} (1969) 493.

\bibitem{Cr51} P. Anselmann et al. {\it Phys. Lett.} {\bf B 342} (1995) 440;
J.N. Abdurashitov et al. {\it Phys. Rev. Lett.} {\bf 77} (1996) 4708; W. Hampel et al. {\it Phys. Lett.} {\bf B 420} (1998) 114

\bibitem{SKNIM} S. Fukuda et al. {\it Nuclear Instruments and Methods} {\bf A 501} (2003) 418.

\bibitem{DayNight} A.~J. Baltz and J. Weneser, {\it Phys. Rev. D} {\bf 37} (1988) 3364;
M.~C. Gonzalez-Garcia, C. Pe\~na-Garay and A.~Y. Smirnov {\it Phys. Rev. D}{\bf63} (2001) 113004.

\bibitem{SKMagMom} D. W. Liu et al (2004) hep-ex/040215 v1.

\bibitem{SNONIM} J. Boger et al. {\it Nucl. Inst. Meth.}{\bf A449} (2000) 72.

\bibitem{SNO1} Q.R. Ahmad et al. {\it Phys. Rev. Lett.}{\bf 87}
(2001) 071301.

\bibitem{SNO} Q.R. Ahmad et al. {\it Phys. Rev. Lett.}{\bf 89}
(2002) 011301

\bibitem{SNO3} Q.R. Ahmad et al. {\it Phys. Rev. Lett.}{\bf 89}
(2002) 011302

\bibitem{KAMLAND} E. Eguchi et al. {\it Phys. Rev. Lett.} {\bf 90} (2003) 021802.  

\bibitem{Bahcallsterile} J.N. Bahcall, M.C. Gonzales-Garcia, C. Pena-Garay {\it JHEP} {\bf 0302} (2003) 009.

\bibitem{KamAntinu} K. Eguchi et al. {\it Phys. Rev. Lett.} to be published, hep-ex/0310047.

\bibitem{Maltoni} M. Maltoni, T. Schwetz, M.A. Tortola, J.W.F. Valle {\it Phys. Rev.} {\bf D68} (2003) 113010.

\bibitem{CPT} John N. Bahcall, V. Barger, and Danny Marfatia), {\it Phys. Lett. B}, {\bf 534}, (2002) 120-123, hep-ph/0201211.

\bibitem{PakvasaBarger} I thank S. Pakvasa and V. Barger for discussions on this topic.

\bibitem{SmirnovSterile} P.C. Holanda, A. Yu. Smirnov, hep-ph/0307266.

\bibitem{BXAP16} Borexino collaboration, G. Alimonti {\it et al.},
	{\it Astropart. Phys. } {\bf 16} (2002) 205.

\bibitem{BXAP18} Borexino collaboration, C. Arpesella {\it et al.}
	{\it Astropart. Phys. } {\bf 18} (2002) 1.

\bibitem{ctf} Borexino collaboration, G. Alimonti {\it et al.},
	{\it Astropart. Phys. } {\bf 8} (1998) 141; 
	Borexino collaboration, G. Alimonti {\it et al.},
	{\it Nucl. Instrum. Meth. }{\bf A406} (1998) 411. 

\bibitem{Yuksel} A. B. Balantekin, V. Barger,  D. Marfatia, S. Pakvasa, H. Yuksel, private communication.

\bibitem{Furuno} K. Furuno for the KamLAND collaboration, LoNU2002...

\bibitem{Schoenert} D. Franco, G. Rannucci, S. Schoenert, private communication for the Borexino collaboration.

\bibitem{xmass}Y. Suzuki, LowNu2 workshop, Tokyo, 2000, 
                \verb|http://www-sk.icrr.u-tokyo.ac.jp/neutlowe/|,
                
\bibitem{heron} B. Lanou, LowNu 2002 workshop, Heidelberg, 2002,
		\verb|http://www.mpi-hd.mpg.de/nubis/www_lownu2002/|.

\bibitem{clean} D.N. McKinsey and J.M. Doyle, {\it J. of Low Temp. Phys.}
  		{\bf 118} (2000) 153.

\bibitem{tpc} G. Bonvicini et al., Contributed paper, Snowmass 2001,
                 hep-ex/0109199, hep-ex/0109032
\bibitem{lens} R.S. Raghavan, {\it Phys. Rev. Lett. }{\bf 78} (1997) 3618;
LENS collaboration, Letter of Intent, Laboratori Nazionali del 
Gran Sasso (1999); \verb|http://www.mpi-hd.mpg.de/nubis/www_lownu2002/|

\bibitem{RaghavanNoon} R. S. Raghavan NOON2004 Conference, Tokyo, Japan, 2004; \verb|http://www-sk.icrr.u-tokyo.ac.jp/noon2004/|
               
\bibitem{moon} H. Ejiri et al., {\it Phys. Rev. Lett. }{\bf 85} (2000) 2917.

\bibitem{C14} Borexino Collaboration, G. Alimonti {\it et al.},
	{\it Phys. Lett. }{\bf B 422} (1998) 349.

\bibitem{Namba} T. Namba NOON2004 Conference, Tokyo, Japan, 2004; \verb|http://www-sk.icrr.u-tokyo.ac.jp/noon2004/|

\bibitem{Nakahata} M. Nakahata, LowNu 2002 Workshop, Heidelberg, 2002,
		\verb|http://www.mpi-hd.mpg.de/nubis/www_lownu2002/|.
\end{thebibliography}
\end{document}